\documentclass{kluwer}    
\newdisplay{guess}{Conjecture}
\def\s|{\left|\!\left|}
\def\d|{\right|\!\right|}

\newcommand{\be}{\begin{equation}}
\newcommand{\ee}{\end{equation}}
\def\erre{\hbox{\rm\rlap{I}\kern.1em R}}

\begin{document}
\begin{article}
\begin{opening}

\title{ A geometric interpretation of integrable motions}
\author{Cecilia \surname{Clementi}\thanks{Electronic address: 
cclementi@ucsd.edu}}
\institute{Department of Physics, University of California at San Diego, 
La Jolla, CA 92093-0319, USA}
\author{Marco \surname{Pettini}\thanks{Also I.N.F.M., Unit\`a di Firenze, and
I.N.F.N., Sezione di Firenze -
Electronic address: pettini@arcetri.astro.it}}
\institute{Osservatorio Astrofisico di Arcetri, 
Largo E.Fermi, 5 - 50125 Firenze, Italy}
\runningauthor{Cecilia Clementi \&\ Marco Pettini}
\runningtitle{Geometric interpretation of integrable motions}

\date {\today}

\begin{abstract}
Integrability, one of the classic issues in galactic dynamics and in general 
in celestial mechanics, is here revisited in a Riemannian geometric
framework, where newtonian motions are seen as geodesics of suitable 
``mechanical'' manifolds. The existence of constants of motion 
that entail integrability is associated with the existence of Killing tensor
fields on the mechanical manifolds. Such tensor fields correspond to hidden
symmetries of non-Noetherian kind. Explicit expressions for Killing tensor 
fields are given for the $N=2$ Toda model, and for a modified H\'enon-Heiles 
model, recovering the already known analytic expressions of the
second conserved quantity besides energy for each model respectively.
\end{abstract}

\end{opening}

\section{Introduction}
The problem of integrability in classical mechanics has been very seminal.
Motivated by celestial mechanics, it has stimulated a wealth of analytical
methods and results. For example, the weaker request of only approximate 
integrability over finite times, or the existence of integrable regions in 
the phase space of a globally non-integrable system, have led to the 
development of classical perturbation theory, with all its important 
achievements.
However, deciding whether a given Hamiltonian system is globally
integrable or not
still remains a difficult task, for which a general constructive framework
is lacking. Besides its theoretical interest, the problem of integrability is
still relevant to a number of open problems among which we can mention a long
standing one in galactic dynamics: the quest for the third integral of
motion besides energy and angular momentum \cite{binney}. 
In fact, the apparent absence
of dynamical chaos in several models describing the motions of test stars 
in mean-field  galactic gravitational potentials suggests that these models
might be integrable. 

The aim of the present paper is to draw attention to the rephrasing of this 
classical problem of integrability in a Riemannian geometric language. Such a
possibility exists because  Hamiltonian flows can be identified with 
geodesic flows on Riemannian manifolds equipped with suitable metrics.
The Riemannian geometric framework has hitherto proved very useful to tackle 
Hamiltonian
chaos \cite{physrep}, to understand the origin of chaos in non-Anosov flows
\cite{Marco,marcomoni,franzo,marcorick,lapcecmar} 
and also to analytically compute Lyapunov exponents \cite{lapcecmar}.

The existence of conservation laws, and of conserved quantities along the
trajectories of a Hamiltonian system, is related with the existence of 
symmetries. The link is made by Noether theorem \cite{arnold}. A symmetry is
seen as an invariance under the action of a group of transformations, and, in
the case of continuous symmetries, this can be related also to the existence
of special vector fields, Killing vector fields on the mechanical manifold, 
generating the transformations. However, through Noetherian symmetries, and
thus Killing vector fields, only a limited set of conservation laws can be
accounted for. This is easily understood because only invariants that are
linear functions of the momenta can be constructed by means of Killing vectors,
while the energy, an invariant for any autonomous Hamiltonian system, is 
already a quadratic function of the momenta.
The possibility of constructing invariants along a geodesic flow, that are of 
higher order than linear in the momenta, is related with the existence of 
Killing {\it tensor} fields on the mechanical manifolds \cite{eisenhart}.
In the present paper we discuss all these facts and we show how it is possible
to explicitly work out the components of the Killing tensors associated with 
two integrable models: an $N=2$ Toda model and a modified version of the
H\'enon-Heiles model, and hence to obtain the analytic expressions of the
second integral of motion besides energy.

In general, the components of any Killing tensor field on a mechanical manifold
are solutions of a linear, non-homogeneous system of first order partial
differential equations. As the number of these equations always exceeds the 
number of the unknowns, the system is always {\it overdetermined}. 
The existence of Killing tensors thus requires {\it compatibility}. However,  
compatibility is generically very unusual, hence a possible explanation, 
at least of qualitative kind, of the exceptionality of
integrability with respect to non-integrability.

For the sake of clarity and self-containedness, we briefly recall some basic 
points about the geometrization of newtonian mechanics, about Killing vector
fields and Killing tensor fields in Sections II, III, and IV, respectively.
Section V contains the original results mentioned above about the
relationship between integrability and the existence of Killing tensor fields
on the mechanical manifolds.

\section{Geometric formulation of Hamiltonian dynamics}
Let us briefly recall a few basic points about the geometrization of
newtonian dynamics in a Riemannian geometric framework. This applies to
dynamical systems described by standard Hamiltonians, i.e.
$H(p,q)=\frac{1}{2}a^{ij}(q)p_ip_j + V(q)$, with the shorthands 
$p=(p_1,\dots ,p_N)$ and $q=(q_1,\dots ,q_N)$. Equivalently, we can
describe these systems through Lagrangian functions
$L( q,\dot{q})= \frac{1}{2}a_{ij}(q)\dot{q}^{i}\dot{q}^{j}- V( q)$.

According to Maupertuis' principle of stationary action, among all the 
possible isoenergetic paths $\gamma (t)$ with fixed end points,
the paths that make vanish the first variation of the action functional
\begin{equation}
{\cal A} = \int_{\gamma(t)} p_i \, dq_i
         = \int_{\gamma(t)} \frac{\partial L}{\partial \dot q_i}
          \, \dot q_i\, dt
\label{action_M}
\end{equation}
are natural motions.

The kinetic energy $W$ is a homogeneous function of degree two, hence 
$2W = \dot q_i \partial L/{\partial \dot q_i}~,$ and Maupertuis'
principle reads
\begin{equation}
\delta{\cal A} = \delta \int_{\gamma(t)} 2W \, dt = 0~.
\label{M_principle}
\end{equation}
The configuration space $M$ of a system with $N$ degrees of freedom is an
$N$-dimensional differentiable manifold and the lagrangian coordinates
$(q_1,\ldots,q_N)$ can be used as local coordinates on $M$. The manifold $M$
is naturally given a proper Riemannian structure. In fact, by introducing  
the matrix
\begin{equation}
g_{ij} = 2[E - V(q)] a_{ij}
\label{gij}
\end{equation}
Eq.(\ref{M_principle}) becomes
\begin{equation}
\delta \int_{\gamma(t)} 2W \, dt =
\delta \int_{\gamma(t)} \left( g_{ij} \dot q^i \dot q^j \right)^{1/2} \, dt =
\delta \int_{\gamma(s)} ds\, = 0 ~,
\end{equation}
so that the newtonian motions fulfil the geodesic condition on the manifold 
$M$, provided we define $ds$ as its arclength. The metric tensor $g_J$ of $M$
is defined through its components by Eq.(\ref{gij}). This is known as Jacobi 
(or kinetic energy) metric.
Denoting by $\nabla$ the canonical Levi-Civita connection on $(M,g_{_J})$, 
the geodesic equation
\begin{equation}
\nabla_{\dot\gamma} \dot\gamma = 0
\end{equation}
becomes, in the local coordinates $(q^1,\ldots,q^N)$,
\begin{equation}
\frac{d^2 q^i}{ds^2} + \Gamma^i_{jk} \frac{dq^j}{ds} \frac{dq^k}{ds} = 0~,
\label{eq_geodesics_loc}
\end{equation}
where the Christoffel coefficients $\Gamma^i_{jk}$ are the components of 
$\nabla$ defined by
\begin{equation}
\Gamma^i_{jk} =  \frac{1}{2} g^{im}
\left( \partial_j g_{km} + \partial_k g_{mj} - \partial_m g_{jk} \right)~
\label{Gamma}
\end{equation}
\[
= -\frac{1}{2 W}[\delta^{i}_{k}\partial _{j}V
+\delta^{i}_{j}\partial_{k}V - \partial_{l}V a^{li}a_{jk}]
+\frac{1}{2}a^{il}[\partial_{l}a_{jk} +
\partial_{k}a_{lj} - \partial_{l}a_{jk}]~,
\]
where $\partial_i = \partial/\partial q^i$. Without loss of generality
consider $g_{ij} = 2[E-V(q)]\delta_{ij}$, so that
\begin{equation}
\Gamma^{i}_{jk}=
-\frac{1}{2 W}[\delta^{i}_{k}\partial _{j}V +\delta^{i}_{j}\partial_{k}V -
\delta_{jk}\partial^{i}V]~.
\label{Connessioni.Jacobi}
\end{equation}
and from Eq. (\ref{eq_geodesics_loc}) we get
\begin{equation}
\frac{d^2 q^i}{ds^2} + \frac{1}{2(E - V)}
\left[2 \frac{\partial (E-V)}{\partial q_j} \frac{dq^j}{ds} \frac{dq^i}{ds}
- g^{ij} \frac{\partial (E-V)}{\partial q_j}
g_{km}\frac{dq^k}{ds} \frac{dq^m}{ds} \right] = 0~,
\end{equation}
and, using $ds^2 = 2(E-V)^2\, dt^2$, these equations finally yield
\begin{equation}
\frac{d^2 q^i}{dt^2} = - \frac{\partial V}{\partial q_i}~,~~~~~~~~~~~
i=1,\dots ,N~.
\end{equation}
which are Newton equations.

\section{Killing vector fields}
On a Riemannian manifold, for any pair of vectors $V$ and $W$, the following
relation holds
\begin{equation}
\frac{d}{ds}\langle V,W\rangle = \langle\frac{\nabla V}{ds},W\rangle +
\langle V,\frac{\nabla W}{ds}\rangle
\label{cov_distr}
\end{equation}
where $\langle V,W\rangle = g_{ij}V^iW^j$ and $\nabla/ds$ is the covariant 
derivative along a curve $\gamma (s)$. If the curve $\gamma (s)$ is a 
geodesic, for a generic vector $X$ we have
\begin{equation}
\frac{d}{ds}\langle X,{\dot\gamma}\rangle = \langle\frac{\nabla X}{ds},
{\dot\gamma}\rangle +
\langle X,\frac{\nabla{\dot\gamma}}{ds}\rangle = \langle\frac{\nabla X}{ds},
{\dot\gamma}\rangle\equiv \langle{\nabla_{\dot\gamma} X}, {\dot\gamma}\rangle
\label{geo_distr}
\end{equation}
where 
$(\nabla_{\dot\gamma}X)^i=\frac{dx^l}{ds}\frac{\partial X^i}
{\partial x^l}+\Gamma^i_{jk}\frac{dx^j}{ds}X^k$, so that in components 
it reads
\begin{equation}
\frac{d}{ds}(X_i v^i) =v^i \nabla_i (X_j v^j)
\label{cons1}
\end{equation}
where $v^i=dx^i/ds$; with 
$X_j v^i\nabla_i v^j=X_j \nabla_{\dot\gamma}\dot\gamma^j=0$ -- because 
geodesics are autoparallel -- this can be obviously rewritten as 
\begin{equation}
\frac{d}{ds}(X_i v^i) = \frac{1}{2} v^jv^i (\nabla_i X_j + \nabla_j X_i)
\label{cons2}
\end{equation}
telling that the vanishing of the l.h.s., i.e. the conservation of $X_i v^i$
along a geodesic, is guaranteed by the vanishing of the
r.h.s., i.e.   
\begin{equation}
\nabla_{(i}X_{j)}\equiv\nabla_{i}X_{j}+\nabla_{j}X_{i}=  0~,~~~~~~ 
i,j=1,\dots ,dim M_E .
\label{kilvec}
\end{equation}
If such a field exists on a manifold, it is called a Killing vector field 
(KVF).
Equation (\ref{kilvec}) is equivalent to ${\cal L}_X g =0$, where ${\cal L}$
is the Lie derivative.
On the mechanical manifolds $(M_E,g_J)$, being the unit vector 
$\frac{d q^{k}}{d s}$ -- tangent to a geodesic -- proportional to the 
canonical momentum $p_{k}=\frac{\partial L}{\partial
\dot{q}^{k}}=\dot{q}^{k},\;(a_{ij}=\delta_{ij})$, the existence
of a KVF $X$ implies that the quantity, linear in the momenta,  
\begin{equation}
J({ q,p})=X_{k}({ q})\frac{dq^{k}}{d s}
=\frac{1}{\sqrt{2}(E-V({q}))}X_{k}( q)\frac{dq^{k}}
{d t} =\frac{1}{\sqrt{2}W( q)}\sum_{k=1}^N X_{k}(q)p_{k}
\label{satura.rango1}
\end{equation}
is a constant of motion along the geodesic flow.
Thus, for an $N$ degrees of freedom Hamiltonian system, a physical 
conservation law, involving a conserved quantity linear in the canonical 
momenta, can always be related with a symmetry on the manifold $(M_E,g_{J})$ 
due to the action of a KVF on the manifold.
These are conservation laws of Noetherian kind.
The equation (\ref{kilvec}) is
equivalent to the vanishing of the Poisson brackets
\begin{equation}
\{ H,J\}=\sum_{i=1}^{N} \left(
\frac{\partial H}{\partial
q^{i}}\frac{\partial J}{\partial p_{i}}-
\frac{\partial H}{\partial
p_{i}}\frac{\partial J}{\partial q^{i}}\right) = 0~,
\label{Parent.Poiss}
\end{equation}
the standard definition of a constant of motion $J( q,p)$.
In fact, a linear function of the momenta 
\begin{equation}
J({q,p})=\sum_{i} C_{i}({q})p_{i}~,
\label{linearJ}
\end{equation}
if conserved, can be associated with the vector of components
\begin{equation}
X_{k}=[E- V(q)] C_{k}({ q}).
\end{equation}
The explicit expression of the system of equations
(\ref{kilvec}) is obtained by writing in components the covariant
derivatives associated with the connection coefficients 
(\ref{Connessioni.Jacobi}) and it finally reads
\begin{equation}
[E-V(q)]
\left[\frac{\partial C_{i}({ q})}{\partial
q^{j}} +
\frac{\partial C_{j}({ q})}{\partial
q^{i}}\right] - \delta_{ij}\sum_{k=1}^{N}\frac{\partial
V}{\partial q^{k}} C_{k}({ q})=0,
\end{equation} 
or equivalently
\begin{equation}
\frac{1}{2} \sum_{k=1}^{N}p_{k}^{2}
\left[\frac{\partial C_{i}({ q})}{\partial
q^{j}} +
\frac{\partial C_{j}({ q})}{\partial
q^{i}}\right]
- \delta_{ij}
\sum_{k=1}^{N}\frac{\partial
V}{\partial q^{k}} C_{k}({ q})=0,
\end{equation}
which, according to the principle of polynomial identity, yields the
following conditions for the coefficients  $C_i({ q})$
\[
\frac{\partial C_{i}(q)}{\partial
q^{j}} +
\frac{\partial C_{j}( q)}{\partial
q^{i}}=0\;\;\;\;\;  i\neq j~,~~i,j=1,\dots,N
\]
\be
\frac{\partial C_{i}(q)}{\partial
q^{i}}=0\;\;\;\;\;  i=1,\dots,N
\label{coeff.rango1}
\ee
\[
\sum_{k=1}^{N}\frac{\partial
V}{\partial q^{k}} C_{k}({q})=0~~.
\]
One can easily check that the same conditions stem from 
Eq.(\ref{Parent.Poiss}).
As an elementary example, we can give the explicit expression of the
components of the 
Killing vector field associated with the conservation of the total momentum
$P(q,p)= \sum_{k=1}^{N}p_{k}~~$.

In this case the coefficients are $C_i({ q})= 1$, so that the 
momentum conservation can be geometrically related with the action of the
vector field of components
\be
X_{i}= E- V(q),\;\;\;\;\;  i=1,\dots,N
\label{tensoreK.disaccopp}
\ee
on the mechanical manifold.
At least this class of invariants has a geometric counterpart in a symmetry
of $(M_E,g_J)$.

However, in order to achieve a fully geometric rephrasing of integrability,
we need something similar for any constant of motion. If a one-to-one
correspondence is to exist between conserved physical quantities along a
Hamiltonian flow and suitable symmetries of the mechanical manifolds 
$(M_E,g_J)$,
then {\it integrability} will be equivalent to the existence of a number of 
symmetries at least equal to the number of degrees of freedom 
($= dim\ M_E$).

If a Lie group $G$ acts on the phase space manifold through completely
canonical transformations, and there exists an associated 
{\it momentum mapping}\footnote{This happens whenever this action corresponds
to the lifting to the phase space of the action of a Lie group on the
configuration space.}, then every Hamiltonian having $G$ as a symmetry group, 
with respect to its action, admits the momentum mapping as constant of motion 
\cite{abram}.
These symmetries are usually referred to as {\it hidden symmetries} because,
even though their existence is ensured by integrability, they are not easily
recognizable\footnote{An interesting account of these hidden symmetries can 
be found in \cite{perelomov} where it is surmised that integrable motions of
$N$ degrees of freedom systems are the ``shadows'' of free motions in
symmetric spaces (for example euclidean spaces ${\mathbb R}^n$, hyperspheres
${\mathbb S}^n$, hyperbolic spaces ${\mathbb H}^n$) of sufficiently large 
dimension $n>N$.}.

\section{Killing tensor fields}
Let us now extend what has been presented in the previous section about KVFs,
trying to generalize the form of the conserved quantity along a geodesic flow
from $J=X_i v^i$ to $J=K_{j_1j_2\dots j_r}v^{j_1}v^{j_2}\dots v^{j_r}$, with 
$K_{j_1j_2\dots j_r}$ a tensor of rank $r$.
Thus, we look for the conditions that entail
\begin{equation}
\frac{d}{ds}(K_{j_1 j_2\dots j_r}v^{j_1}v^{j_2}\dots v^{j_r}) =
v^j\nabla_j (K_{j_1j_2\dots j_r}v^{j_1}v^{j_2}\dots v^{j_r}) = 0~.
\label{higher-cons}
\end{equation}
In order to work out from this equation a condition for the existence of a
suitable tensor $K_{j_1j_2\dots j_r}$, which is called a {\it Killing 
tensor field} (KTF), let us first consider the $2r$ rank tensor 
$K_{j_1j_2\dots j_r}v^{i_1}v^{i_2}\dots v^{i_r}$ and its covariant derivative 
along a geodesic, i.e.
\[
v^j\nabla_j (K_{j_1j_2\dots j_r}v^{i_1}v^{i_2}\dots v^{i_r})=
\]
\[
= v^j\left(\frac{\partial K_{j_1\dots j_r}}{\partial x^j} - K_{lj_2\dots j_r}
\Gamma^l_{j_1j}- \dots - K_{j_1\dots l}\Gamma^l_{j_rj}\right)
v^{i_1}\dots v^{i_r}\ +
\]
\[
+K_{j_1\dots j_r}\left( v^j\frac{\partial v^{i_1}}{\partial x^j} + 
\Gamma^{i_1}_{jl} v^l v^j\right)v^{i_2}\dots v^{i_r}+
\dots 
\]
\[
\ldots +K_{j_1\dots j_r}v^{i_1}\dots v^{i_{r-1}}
\left( v^j\frac{\partial v^{i_r}}{\partial x^j} + 
\Gamma^{i_r}_{jl} v^l v^j\right)
\]
\begin{equation}
=v^{i_1}v^{i_2}\dots v^{i_r} v^j\nabla_j K_{j_1j_2\dots j_r}
\label{higher-cons1}
\end{equation}
where we have again used  $v^j\nabla_j v^{i_k}=0$ along a geodesic, and a 
standard covariant differentiation formula \cite{novikov}.
Now, by contraction on the indices $i_k$ and $j_k$ the $2r$-rank tensor of 
the r.h.s. of Eq.(\ref{higher-cons1}) provides a new expression 
for the r.h.s. of Eq.(\ref{higher-cons}) which reads
\begin{equation}
\frac{d}{ds}(K_{j_1 j_2\dots j_r}v^{j_1}v^{j_2}\dots v^{j_r}) =
v^{j_1}v^{j_2}\dots v^{j_r} v^j\nabla_{(j}K_{j_1 j_2\ldots j_r)}~~,
\label{KTF-eq}
\end{equation}
where $\nabla_{(j}K_{j_1 j_2\ldots j_r)}=\nabla_jK_{j_1 j_2\dots j_r} +
\nabla_{j_1}K_{j j_2\dots j_r}+\dots +\nabla_{j_r}K_{j_1 j_2\dots j_{r-1}j}$,
as it can be easily understood by rearranging the indices of the summations
in the contraction of the $2r$-rank tensor in the last part of Eq.
(\ref{higher-cons1}); (a direct check for the case $N=r=2$ is immediate).
The vanishing of Eq.(\ref{KTF-eq}), entailing the conservation of
$K_{j_1 j_2\dots j_r}v^{j_1}v^{j_2}\dots v^{j_r}$ along a geodesic flow,
is therefore guaranteed by the existence of a
tensor field fulfilling the conditions 
\be
\nabla_{(j}K_{j_1 j_2\ldots j_r)}=0,
\label{eq.tens.Kill}
\ee
these equations generalize Eq.(\ref{kilvec}) and give the definition
of a KTF on a Riemannian manifold. These $N^{r+1}$ equations in 
$(N+r-1)! /r! (N-1)!$ unknown independent components\footnote{This number of 
independent components, i.e. the binomial coefficient ${N+r-1\choose r}$, 
is due to the totally symmetric character of Killing tensors.} 
of the Killing tensor 
constitute an {\it overdetermined} system of equations. Thus, a-priori, we 
can expect that the existence of KTFs has to be rather exceptional.

If a KTF exists on a Riemannian manifold, then the scalar
\be
K_{j_{1} j_{2}\ldots j_{r}} 
\frac{dq^{j_{1}}}{d s}
\frac{dq^{j_{2}}}{d s}\dots
\frac{dq^{j_{r}}}{d s}
\ee 
is a constant of motion for the geodesic flow on the same manifold.

Let us consider, as a generalization of the special case of rank one
given by Eq.(\ref{linearJ}), the following constant of motion
\be
J({ q,p})=\sum_{\{i_{1},i_{2},\ldots,i_{N}\}} C_{i_{1} i_{2}\ldots
i_{N}} p_{1}^{i_{1}}p_{2}^{i_{2}}\ldots p_{N}^{i_{N}},
\label{inv.rango.r}
\ee
which, with the constraint $i_1 + i_2 + \ldots + i_N = r$, is a homogeneous 
polynomial of degree $r$. The index $i_j$ denotes the power with which the
momentum $p_j$ contributes. If $r < N$ then necessarily some indices
$i_j$ must vanish. By repeating the procedure developed in the case $r=1$,
and by identifying
\be
J({q,p})\equiv K_{j_{1} j_{2}\ldots j_{r}} 
\frac{dq^{j_{1}}}{d s}
\frac{dq^{j_{2}}}{d s}\dots
\frac{dq^{j_{r}}}{d s}
\ee 
we get the relationship between the components of the Killing tensor of rank 
$r$ and the coefficients $C_{i_{1} i_{2}\ldots i_{N}}$ of the invariant
$J({ q,p})$, that is
\be
K_{\underbrace{1\dots 1}_{i_{1}},\underbrace{2\ldots
2}_{i_{2}},\ldots,\underbrace{N\ldots N}_{i_{N}}}= 2^{r/2}[E-V({ q})]^r 
C_{i_{1} i_{2}\ldots i_{N}}~.
\ee
With the only difference of a more tedious combinatorics, also in this case it
turns out that the equations (\ref{eq.tens.Kill}) are equivalent to the
vanishing of the Poisson brackets of $J({ q,p})$, that is
\be
\{ H, J\}=0 \Longleftrightarrow \nabla_{(j}K_{j_1 j_2\ldots j_r)}=0~.
\ee
Thus, the existence of Killing tensor fields, obeying 
Eq.(\ref{eq.tens.Kill}), 
on a mechanical manifold $(M,g_{J})$ provide the rephrasing of integrability 
of Newtonian equations of motion or, equivalently, of standard Hamiltonian 
systems, within the Riemannian geometric framework .

At first sight, it might appear too restrictive that prime integrals of motion
have to be homogeneous functions of the components of $ p$. However, as
we shall discuss in the next Section, the integrals of motion of the known
integrable systems can be actually cast in this form. This is in particular 
the case of
total energy, a quantity conserved by any autonomous Hamiltonian system.

\section{Explicit KTFs of known integrable systems}
The first natural question to address concerns the existence of a KT
field, on any mechanical manifold $(M,g_J)$, to be associated with 
total energy conservation. Such a KT field actually exists and coincides 
with the metric tensor $g_J$, in fact it satisfies\footnote{A property of
the canonical Levi-Civita connection, on which the covariant derivative is 
based, is just the vanishing of $\nabla g$.} by definition
Eq.(\ref{eq.tens.Kill}).

One of the simplest case of integrable system is represented by a decoupled
system described by a generic Hamiltonian
\be
H=\sum_{i=1}^{N}\left[ \frac{p_{i}^{2}}{2}+ V_{i}(q_{i})\right] =
\sum_{i=1}^{N} H_{i}(q_{i},p_{i})
\ee
for which all the energies $E_i$ of the subsystems  
$H_i,\; i=1,\ldots,N$, are conserved. On the associated mechanical 
manifold, $N$ KT fields of rank $2$ exist, they are given by
\be
K^{(i)}_{jk}= \delta_{jk} \{ V_{i}(q_{i})[E-V({ q})]+\delta^{i}_{j}
[E- V({q})]^2\}~.
\label{sistemi.disaccopp}
\ee
In fact, these tensor fields fulfil Eq.(\ref{eq.tens.Kill}) which
explicitly reads
\[
\nabla_{k}K^{(i)}_{lm}+
\nabla_{l}K^{(i)}_{mk}+
\nabla_{m}K^{(i)}_{kl}=
\]
\be
=\frac{\partial K^{(i)}_{lm}}{\partial q^{k}}+
\frac{\partial K^{(i)}_{mk}}{\partial q^{l}}+
\frac{\partial K^{(i)}_{kl}}{\partial q^{m}}
-2\Gamma^{j}_{kl}K^{(i)}_{jm}-
2\Gamma^{j}_{km}K^{(i)}_{jl}-
2\Gamma^{j}_{lm}K^{(i)}_{jk} = 0
\ee
\[
 k,l,m = 1,\ldots,N~~. 
\]
The conserved quantities $J^{(i)}({ q,p})$ are then obtained by 
saturation of the tensors $K^{(i)}$ with the velocities $d{ q}/ds$
\be
J^{(i)}({ q,p})=
\sum_{jk=1}^{N}K^{(i)}_{jk}\frac{d q^{j}}{d s}\frac{d q^{k}}{d s}=
V_{i}(q_{i})\frac{1}{
E-V({ q})}\sum_{k=1}^{N}\frac{p_{k}^{2}}{2}+ \frac{p_{i}^{2}}{2}=
E_{i}~.
\ee
This equation suggests that to require that the constants of motion have to
be homogeneous polynomials of the momenta is not so restrictive as it might 
appear, in fact, through the following constant quantity 
\be
\frac{1}{E-V({ q})}\sum_{k=1}^{N} \frac{p_{k}^2}{2}=1
\ee
homogeneous of second degree in the momenta, 
any even degree polynomial of the momenta can be made homogeneous. 
The possibility of inferring the existence of a conservation law from the 
existence of a KTF on $(M,g_J)$ is thus extended to the constants 
of motion given by a sum of homogeneous polynomials whose degrees differ by
an  even integer
\be
J({ p,q})= 
P^{(r)}({ p})+ P^{(r-2)}({ p})+
\ldots +
\ee
\[
+P^{(r-2n)}({ p})\in C^{\infty}({ q})
[p_{1},\ldots,p_{N}]
\]
\[
homdeg P^{s}=s\;\;\;\; s=r,r-2,\ldots,r-2[\frac{r}{2}]
\]
so that it can be recast in the homogeneous form
\be
J({ p,q})= 
P^{(r)}({ p})+P^{(r-2)}({ p})
\frac{1}{E-V({ q})}\sum_{k=1}^{N} \frac{p_{k}^2}{2}+
\ldots +
\ee
\[
+P^{(r-2n)}({ p})
\left[\frac{1}{E-V({ q})}\sum_{k=1}^{N} \frac{p_{k}^2}{2}\right]^n~.
\]
\subsection{Nontrivial integrable models}
Nontrivial examples of nonlinear integrable Hamiltonian systems are provided
by the following Hamiltonians  
\begin{equation}
H=\sum_{i=1}^{N}\left\{\frac{p_{i}^{2}}{2} +
\frac{a}{b}[e^{-b(q_{i+1}-q_{i})}-1]\right\}
\label{todaN}
\end{equation}
known as the Toda model \cite{Toda}, which is integrable for any given pair 
of the constants $a$ and $b$, and
\be
H=\sum_{i=1}^{N}\frac{p_{i}^{2}}{2} + 
\frac{1}{2}\left(\sum_{i=1}^{N}q_i^2\right)^2 - \sum_{i=1}^{N}\lambda_iq_i^2
\ee
which is completely integrable for any $\lambda_1,\dots ,\lambda_N$
\cite{choodnovsky}.
Recursive formulae are available for all the constants of motion of the Toda
model at any $N$ \cite{Henon_Toda}, and also for the second Hamiltonian the 
exact form of first integrals is known \cite{choodnovsky}. In both cases, the
first integrals are polynomials of given parity of the momenta so that, 
on the basis of what we have said above, each invariant $J^{(i)},~i=1,\dots,N$
can be derived from a KTF on $(M,g_{J})$. Thus, integrability of these
systems admits a Riemannian-geometric interpretation.

\subsection{The special case of N=2 Toda model}
Let us consider the special case of a two-degrees of freedom Toda model
described by  the integrable Hamiltonian\footnote{This is derived from an $N=3$
Hamiltonian (\ref{todaN}) by means of two canonical transformations of 
variables removing translational invariance, see for example 
\cite {boccaletti}; the third order expansion of this new Hamiltonian yields
the H\'enon-Heiles model of Eq.(\ref{HH_CD}) with $C=D=1$.}
\be
H=\frac{1}{2}(p_{x}^{2}+p_{y}^{2}) + \frac{1}{24}\left[ e^{2y+2\sqrt{3}x}
+ e^{2y-2\sqrt{3}x} + e^{-4y}\right] - \frac{1}{8}~.
\label{Toda2}
\ee
From what is already reported in the literature \cite{Henon_Toda}, we know 
that a third order polynomial of the momenta has to be found eventually, 
therefore we look for a rank-$3$ KT fulfilling
\begin{equation}
\nabla_i K_{jkl} + \nabla_j K_{ikl}+ \nabla_k K_{ijl}+ \nabla_l K_{ijk}=0~,
~~~~~~~i,j,k,l=1,2
\label{TodaKT}
\end{equation}
where, associating the label $1$ to $x$ and the label $2$ to $y$, 
$\{(i,j,k,l)\}=\{(1,1,1,1);(1,1,1,2);(1,1,2,2);(1,2,2,2);(2,2,2,2)\}$.
The computation of the Christoffel coefficients according to 
Eq.(\ref{Connessioni.Jacobi}) yields
\[
\Gamma^1_{11}=\frac{-\partial_xV}{2[E-V(x,y)]}~,~~~
\Gamma^1_{22}=\frac{\partial_xV}{2[E-V(x,y)]}~,~~~
\Gamma^2_{11}=\frac{\partial_yV}{2[E-V(x,y)]}~,~~~
\]
\begin{equation}
\Gamma^2_{22}=\frac{-\partial_yV}{2[E-V(x,y)]}~,~~~
\Gamma^1_{12}=\frac{-\partial_yV}{2[E-V(x,y)]}~,~~~
\Gamma^2_{12}=\frac{-\partial_xV}{2[E-V(x,y)]}~.\\
\label{TodaChr}
\end{equation}
From Eq.(\ref{TodaKT}) we get the system
\begin{eqnarray}
\nabla_1K_{111}&=& 0 \nonumber\\
\nabla_1K_{122} + \nabla_2K_{112}&=& 0 \nonumber\\
\nabla_2K_{111} + 3\nabla_1K_{211} &=& 0 \nonumber\\
\nabla_1K_{222} + 3\nabla_2K_{122}&=& 0 \nonumber\\
\nabla_2K_{222}&=& 0 
\label{Todasys}
\end{eqnarray}
whence
\begin{eqnarray}
\partial_xK_{111}-3\Gamma^1_{11}K_{111}-3\Gamma^2_{11}K_{211}&=& 0 \nonumber\\
\partial_xK_{122} + \partial_yK_{211} -\Gamma^1_{11}K_{122}-
\Gamma^2_{11}K_{222}-4\Gamma^1_{12}K_{112}-&&\nonumber\\
4\Gamma^2_{11}K_{212}
-\Gamma^1_{22}K_{111}- \Gamma^2_{22}K_{211}&=& 0 \nonumber\\
\partial_yK_{111} + 3\partial_xK_{211} -6\Gamma^1_{12}K_{111}-
6\Gamma^2_{12}K_{112}-&&\nonumber\\
6\Gamma^1_{11}K_{211}-6\Gamma^2_{11}K_{212}&=& 0 
\nonumber\\
\partial_xK_{222} + 3\partial_yK_{122} -6\Gamma^1_{21}K_{122}-
6\Gamma^2_{21}K_{222}-&&\nonumber\\
6\Gamma^1_{22}K_{112}-6\Gamma^2_{22}K_{212}&=& 0 
\nonumber\\
\partial_yK_{222}-3\Gamma^1_{22}K_{122}-3\Gamma^2_{22}K_{222}&=& 0
\label{Todasys1}
\end{eqnarray}
with the Christoffel coefficients given by Eq.(\ref{TodaChr}), where one has to
replace $V(x,y)$ with the potential part of the Hamiltonian (\ref{Toda2})
and $\partial_xV$, $\partial_yV$ with its derivatives. The general 
method of solving a linear, non-homogeneous system of first-order partial
differential equations in more than one dependent variables is sketched
in Appendix.
However, finding the explicit solution to the system of equations 
(\ref{Todasys1}) is much facilitated because we already know a-priori that
this system is compatible and thus admits a solution, and we also have strong
hints about the solution itself because the general form of the integrals of
the Toda model is known \cite{Henon_Toda}. 
The KTF, besides the metric tensor, for
the model (\ref{Toda2}) is eventually found to have the components
\begin{eqnarray}
K_{111}&=& 2(E-V)^2[ 3\partial_yV +4 (E- V)]\nonumber\\
&=& 8(E-V)^3 +\frac{1}{2}(E-V)^2
[e^{2y-2\sqrt{3}x} +e^{2y+2\sqrt{3}x}-2 e^{-4y}]\nonumber\\
K_{122}&=& 2(E-V)^2[ \partial_yV - 4 (E- V)]\nonumber\\
&=& -24(E-V)^3 +\frac{1}{2}(E-V)^2
[e^{2y-2\sqrt{3}x} +e^{2y+2\sqrt{3}x}-2 e^{-4y}] \nonumber\\
K_{112}&=& -2(E-V)^2 \partial_xV 
= \frac{\sqrt{3}}{6}(E-V)^2
(e^{2y+2\sqrt{3}x} - e^{2y-2\sqrt{3}x})  \nonumber\\
K_{222}&=& -6(E-V)^2 \partial_xV 
=\frac{\sqrt{3}}{2}(E-V)^2
(e^{2y+2\sqrt{3}x} - e^{2y-2\sqrt{3}x}),
\label{KTToda}
\end{eqnarray}
as can be easily checked by substituting them into Eqs.(\ref{Todasys1}). 
Hence, the second constant of motion, besides energy, is given by  
\[
J(x,y,p_x,p_y) = K_{ijk}\frac{dq^i}{ds}\frac{dq^j}{ds}\frac{dq^k}{ds}
= K_{ijk}\frac{dq^i}{dt}\frac{dq^j}{dt}\frac{dq^k}{dt}
\frac{1}{2\sqrt{2}[E-V(x,y)]^3}  
\]
\smallskip
\[
=\frac{1}{2\sqrt{2}[E-V(x,y)]^3} ( K_{111}p_x^3 + 3K_{122}p_x p_y^2 + 
3 K_{112}p_x^2p_y +K_{222}p_y^3)
\]
\smallskip
\[
= 8 p_x (p_x^2 - 3p_y^2) + (p_x +\sqrt{3}p_y) e^{2y-2\sqrt{3}x} - 
2p_x e^{-4y} + (p_x -\sqrt{3}p_y) e^{2y+2\sqrt{3}x}
\]
\begin{equation}
\label{secint}
\end{equation}
which coincides with the expression already reported in the literature 
\cite{lichtenberg} for the Hamiltonian (\ref{Toda2}).

\subsection{The generalized Henon-Heiles model}
Let us now consider the two-degrees of freedom system described by the 
Hamiltonian
\be
H=\frac{1}{2}(p_{x}^{2}+p_{y}^{2}) + \frac{1}{2}(x^{2}+y^{2}) + D x^{2}y
- \frac{1}{3} C y^{3}~.
\label{HH_CD}
\ee
This model, originally derived to describe the motion of a test star in an 
axisymmetric galactic mean gravitational field, provided one of the first 
numerical evidences of the chaotic transition in nonlinear 
Hamiltonian systems  \cite{Henon_Heiles}. H\'enon and Heiles 
considered
the case $C=D=1$. The existence of a chaotic layer in the phase space of this
model means lack of {\it global} integrability. However, by means of the 
Painlev\'e method, it has been shown 
\cite{weiss} that for special choices of the parameters $C$ and $D$
this system is globally integrable. Let us now tackle integrability of this 
model from the viewpoint of the existence of KT fields on the manifold 
$(M,g_{J})$. We first begin with the equations for a Killing vector field.
By means of Eqs.(\ref{coeff.rango1}) we look for possible coefficients 
$C_1(x,y),\; C_2(x,y)$, thus obtaining 
\[
C_{1}=C_{1}(y),\;\;\;\;C_{2}=C_{2}(x)
\]
\be
\frac{d C_{1}(y)}{d y}+ \frac{d C_{2}(x)}{d x}=0
\label{eq.vett.Henon}
\ee
\[
x(1+2Dy)C_{1}(y)+(y+Dx^{2}-Cy^{2})C_{2}(x)=0
\]
From the second equation of Eqs.(\ref{eq.vett.Henon}) it follows that  
\be
\frac{d C_{1}(y)}{d y}=-\frac{d C_{2}(x)}{d x}=cost.
\ee
whence, denoting with $\alpha$ the constant, the possible expressions for 
$C_{1}(y)$ and  $C_{2}(x)$ are only of the form 
$C_{1}(y)=-\alpha y+\beta~,~C_{2}(x)=\alpha x+\gamma$,
that, after substitution into the last equation of Eqs. (\ref{eq.vett.Henon}),
imply 
\be
(x+2Dxy)(-\alpha y+\beta)+(y+Dx^{2}-Cy^{2})(\alpha x+\gamma)=0,
\ee
which has only a non-trivial solution for $C=D=0$. On the other hand, for 
these values of the parameters the potential simplifies to 
$V(x,y)=\frac{1}{2}x^2 + \frac{1}{2}y^{2}$
whence the existence of the Killing vector field $X$ of components $X_1=y$ and
$X_2=-x$ which is due to the invariance under rotations in the $xy$ plane.

Let us now consider the case of a rank-$2$ KTF. 
Equations (\ref{TodaKT}) become   
\begin{equation}
\nabla_i K_{jk} + \nabla_j K_{ik}+ \nabla_k K_{ij}=0~,
~~~~~~~i,j,k=1,2
\label{HH-KT}
\end{equation}
where, associating again the label $1$ to $x$ and the label $2$ to $y$, 
$\{(i,j,k)\}=\{(1,1,1);(1,1,2);(1,2,2);(2,2,2)\}$.
The Christoffel coefficients are still given by Eq.(\ref{TodaChr}), where  
we have to use the potential part of Hamiltonian (\ref{HH_CD}) so that
$\partial_{x}V(x,y)=x+2Dxy$ and $\partial_{y}V(x,y)=y+Dx^{2}-Cy^{2}$.
The KTF equations are then 
\begin{eqnarray}
\nabla_1K_{11}&=& 0 \nonumber\\
2\nabla_1K_{12} + \nabla_2K_{11}&=& 0 \nonumber\\
\nabla_1K_{22} + 2\nabla_2K_{12}&=& 0 \nonumber\\
\nabla_2K_{22}&=& 0 
\label{HH-sys}
\end{eqnarray}
whence
\begin{eqnarray}
\partial_xK_{11}-2\Gamma^1_{11}K_{11}-2\Gamma^2_{11}K_{21}&=& 0 \nonumber\\
2\partial_xK_{12} + \partial_yK_{11} -4\Gamma^1_{12}K_{11}-
(4\Gamma^2_{12}+2\Gamma^1_{11})K_{12}-2\Gamma^2_{11}K_{22}&=& 0 \nonumber\\
\partial_xK_{22} + 2\partial_yK_{12} -2\Gamma^1_{22}K_{11}-
(4\Gamma^1_{12}+2\Gamma^2_{22})K_{12}-4\Gamma^2_{12}K_{22}&=& 0 
\nonumber\\
\partial_yK_{22}-2\Gamma^1_{22}K_{12}-2\Gamma^2_{22}K_{22}&=& 0~.
\nonumber
\end{eqnarray}
\begin{equation}
\label{HH-sys1}
\end{equation}
Since the Hamiltonian (\ref{HH_CD}) is not integrable for a generic choice
of the parameters $C$ and $D$, we can reasonably expect that the generic
property of the above {\it overdetermined} system of equations is
{\it incompatibility}, i.e. only the trivial solution $K_{ij}=0$ exists
for the overwhelming majority of the pairs $(C,D)$. However, the existence
of special choices of $C$ and $D$ for which the Hamiltonian is integrable
suggests that this overdetermined system can be {\it compatible} in special
cases. For example, when $D=0$ the variables $x$ and $y$ in (\ref{HH_CD}) 
are decoupled and thus two KT fields of rank $2$ exist according to 
Eq.(\ref{sistemi.disaccopp}).

A non trivial solution for the system (\ref{HH-sys1}) must exist at least
for the pair $(C=-6,D=1)$. In fact, in this case the modified H\'enon-Heiles
model is known  to be integrable \cite{weiss}.
An explicit solution for the system (\ref{HH-sys1}) is eventually found 
to be given by  
\[
K_{11}=(3-4y)(E-V(x,y))^{2}+x^{2}(x^{2}+4y^{2}+4y+3)(E-V(x,y))
\]
\begin{equation}
K_{12}=2x(E-V(x,y))
\end{equation}
\[
K_{22}=\frac{1}{2}(x^{2}+4y^{2}+4y+3)(E-V(x,y))~.
\]
The associated constant of motion is therefore  
\[
J(x,y,p_{x},p_{y})=\frac{1}{(E-V(x,y))^{2}}(K_{11}p_{x}^{2} + 
2 K_{12}p_{x}p_{y}+ K_{22}p_{y}^{2})=
\]
\smallskip
\begin{equation}
= x^{4}+4x^{2}y^{2}-p_{x}^{2}y+4p_{x}p_{y}x+4x^{2}y +3p_{x}^{2}+3x^{2}~.
\label{secondint}
\end{equation}
This expression is identical to that reported in \cite{weiss}, 
worked out for the same values of $C$ and $D$ with a completely different
method based on the Painlev\'e property.

\section{Concluding remarks}
Let us now summarize the meaning of the results presented above and point
out the open problems.
\smallskip
\begin{itemize}
\item{ } Besides qualitative and quantitative descriptions of chaos, within 
the framework of Riemannian geometrization of newtonian mechanics also
{\it integrability} has its own place. The idea of associating KTFs
with integrability is not new, though this  has been essentially developed 
in the context of classical General Relativity, see for example 
\cite{baleanu,gibbons,sommers} and references quoted therein. Recently, also an
extension to classical newtonian mechanics has been considered in 
\cite{pucacco}, where integrability conditions for 
quadratic invariants were obtained, and where the authors concluded saying
``It is of considerable interest to develop these techniques and use them to 
look for fixed energy invariants of 
physically interesting models such as the H\'enon-Heiles potential and 
others'': this is just what has been done in our present work.
\smallskip
\item{ } The reduction of the problem of integrability of a given 
Hamiltonian system to the existence of suitable KTFs on $(M_E,g_{_{J}})$
offers several reasons of interest, in particular we have seen that the system
of equations in the unknown components of a KTF of a preassigned rank is
overdetermined, thus -- at a qualitative level -- integrability seems a
rather exceptional property, and the larger $N$ the ``more exceptional'' it
seems to be, because of the fastly growing mismatch between the number of
unknowns and the number of equations. In principle, the existence of 
compatibility conditions for systems of linear, first-order partial 
differential equations could allow to decide about integrability prior to any
explicit attempt at solving the equations for the components of a KTF.
Even better, there are geometric constraints to the existence of KTFs, early
results in  this sense are reported in \cite{yano}, so that it seems
possible, at least in some cases, to devise purely geometric criteria of
{\it non-integrability}. 
For example, hyperbolicity of compact manifolds excludes \cite{yano}
the existence of KTFs, and this is consistent with the property of geodesic 
flows on compact hyperbolic manifolds of being strongly chaotic (Anosov 
flows).
\smallskip
\item{ } In the present paper, before working out the second 
invariant besides energy for two integrable models, we already knew 
that a KTF had to exist and of which rank (because of the degree of the 
polynomial invariant), thus we knew that the system of equations to be solved 
was compatible. Whereas, in general we lack a criterion to restrict the
search for  KTFs to a small interval of ranks, and this constitutes a
practical difficulty. Nevertheless, since the involution of two invariants
translates into the vanishing of special brackets -- the Schouten brackets 
\cite{sommers} -- between the corresponding Killing tensors, a shortcut to
prove integrability, for a large class of systems fulfilling the conditions
of the Poincar\'e-Fermi theorem \cite{PF,fermi}, might be to find 
{\it only one} KTF of 
vanishing Schouten brackets with the metric tensor. In fact, for 
quasi-integrable systems with $N\geq 3$, the Poincar\'e-Fermi theorem states
that generically only energy is conserved, thus if another constant of motion
is known to exist (apart from Noetherian ones, like angular momentum) then 
the system must be integrable and in fact there must be $N$ constants 
of motion.
\smallskip
\item{ } At variance with Killing vectors, which are associated with
Noetherian symmetries and conservation laws, Killing tensors no longer have
a simple geometrical interpretation \cite{gibbons,rosquist}, therefore the
associated symmetries are non-Noetherian and hidden.
\end{itemize}
The present paper contributes the subject of a Riemannian geometric approach
to integrability with constructive examples that non-trivial
constants of motion besides energy can be derived from KTFs for two degrees
of freedom integrable Hamiltonians of physical interest. 
This approach to integrability deserves
further attention and investigation. In fact, among the other reasons of
interest, by considering, for example, the standard H\'enon-Heiles model 
($C=D=1$), we could wonder whether the regular regions of phase space
correspond to a 
{\it local} fulfilment of the compatibility conditions of the system 
(\ref{HH-sys1}), this would lead to a better understanding of the relationship
between geometry and stability of newtonian mechanics. Moreover, we could 
imagine that, by suitably defining {\it weak} and {\it strong violations}
of these compatibility conditions, we could better understand the
geometric origin of {\it weak} and {\it strong chaos} in Hamiltonian dynamics
\cite{petlan,petcer} and, perhaps, this might even suggest a starting point to
develop a ``geometric perturbation theory'' complementary to the more standard
canonical perturbation theory.

Finally, the celebrated problem of the third integral in galactic dynamics 
could find here a new constructive, and hopefully useful, approach.

\acknowledgements
The authors are indebted with G. Vezzosi for his friendly and precious help.
It is also a pleasure to thank G. Marmo and A.M. Vinogradov for several
useful discussions and comments.

\appendix

Let us now briefly sketch a classical method \cite{forsyth} of solving 
systems of linear,
first-order, partial differential equations in several dependent variables,
denoted by $z_1,\dots ,z_m$, and two independent variables,
denoted by $x$ and $y$. Writing $X_i=\partial z_i/\partial x$ and 
$Y_i=\partial z_i/\partial y$, equations (\ref{Todasys1}) and (\ref{HH-sys1})
are in the form
\begin{equation}
X_i = B_i + \sum_{s=1}^m A_{is} Y_s
\label{eqlin}
\end{equation}
with an obvious meaning of the coefficients $A_{is}$ and $B_i$; 
$i=1,\dots ,m$. The first step consists of ``diagonalizing'' the above system,
writing equivalent first order equations, or systems of equations, in only
one dependent variable.
Thus equations (\ref{eqlin}) are multiplied by $\lambda_1,\dots ,\lambda_m$,
a set of multipliers, and summed to give
\begin{equation}
\sum_{i=1}^m\lambda_iB_i -\sum_{i=1}^m\lambda_iX_i + \sum_{i,s=1}^m\lambda_i 
A_{is} Y_s = 0~.
\label{sys1}
\end{equation}
Since the derivatives must fulfil the obvious relations
\begin{equation}
d z_i - X_i dx - Y_i dy =0~,
\label{forma}
\end{equation}
by combining Eqs.(\ref{sys1}) and (\ref{forma}) 
the following system of ordinary equations is formed
\begin{equation}
\frac{\sum_{i=1}^m\lambda_i dz_i}{\sum_{i=1}^m\lambda_iB_i}= dx = -
\frac{\lambda_s dy}{\sum_{i=1}^m\lambda_iA_{is}}~,
\label{ode}
\end{equation}
$s=1,\dots,m$, whence, by putting $dy = \mu dx~$, we obtain
\begin{equation}
\sum_{i=1}^m\lambda_iA_{is} + \mu \lambda_s =0
\label{eqcar}
\end{equation}
so that, by solving the critical equation ${\rm det}(A +\mu {\Bbb I})=0$, 
where ${\Bbb I}$ is the identity matrix, 
the numbers
$\lambda_1,\dots ,\lambda_m$ can be eliminated among these equations 
obtaining a set of ratios $\lambda_1:\lambda_2:\dots :\lambda_m$, then
the quantities $\alpha_i$ defined by
\begin{equation}
\alpha_i=\frac{\lambda_i}{\sum_{i=1}^m\lambda_iB_i}
\label{alfai}
\end{equation}
are uniquely determined. The above system of ordinary equations is finally
rewritten as
\begin{eqnarray}
\alpha_1dz_1 +\dots +\alpha_mdz_m & = & dx\nonumber\\
dy & = & \mu dx
\label{auxil}
\end{eqnarray} 
i.e. two linear characteristic equations for each root 
$\mu$ of the critical equation.
Now, if $u(x,y,z_1,\dots,z_m) = const$ is an integral of the equations 
(\ref{auxil}), then it fulfils also the system of $m$ equations
\begin{eqnarray}
\frac{\partial u}{\partial z_1} + \alpha_1 \frac{\partial u}{\partial x} +
\alpha_1\mu\frac{\partial u}{\partial y} &=& 0\nonumber\\
.~.~.~.~.~.~.~.~.~~~~~~~~~~~~~~&&\nonumber\\
\frac{\partial u}{\partial z_m} + \alpha_m \frac{\partial u}{\partial x} +
\alpha_m\mu\frac{\partial u}{\partial y} &=& 0
\label{jacobi}
\end{eqnarray}
since $i=1,\dots,m$, there are $m$ equations in $m+2$ variables; an integral 
of this system, involving any of the dependent variables, is an integral of 
the original system of equations, and if each of the roots of the critical
equation leads to an integral, then the ensemble of these integrals provides
an integral equivalent of the original system.
With  the substitutions
\begin{eqnarray}
z_1& = & w_1\nonumber\\
z_i & = & c_i + (w_1-c_1) w_i~,~~~~~~i=2,\dots,m
\label{sost}
\end{eqnarray}
the following single equation is constructed
\begin{equation}
\frac{\partial u}{\partial w_1} + U_1\frac{\partial u}{\partial x} +
U_2\frac{\partial u}{\partial y} = 0
\label{single}
\end{equation}
where $U_1=\alpha_1+\sum_{i=2}^m\alpha_i w_i$ and $U_2=\alpha_1\mu +
\sum_{i=2}^m \mu \alpha_i w_i$. Finally, the integration of the equation
(\ref{single}) proceeds by integrating the characteristic 
equations 
\begin{equation}
dw_1 = \frac{dx}{U_1} = \frac{dy}{U_2}
\label{auxil1}
\end{equation}
taking $w_2,\dots,w_m$ as non-varying quantities, and then proceeding in a 
standard way \cite{hilbert}.

\end{article}
\end{document}